\documentclass[letterpaper, 10 pt, conference]{ieeeconf}  % Comment this line out if you need a4paper
\usepackage{titlesec}
\usepackage{xspace}

\newcommand{\projectname}{\emph{Surgi-HDTMR}\xspace}
\usepackage{tabularx}   % 自适应列宽表格
\usepackage{booktabs}   % \toprule, \midrule, \bottomrule
\usepackage{array}      % 自定义列类型（如 >{\raggedright\arraybackslash}X）
\usepackage{comment}
\usepackage{tabularx,booktabs,array}

\usepackage{xcolor}
\usepackage{amsmath,amssymb}
\usepackage{mathtools} % 可选: 细节增强(如 \DeclarePairedDelimiter)

\usepackage{graphicx}
\usepackage[caption=false,font=footnotesize]{subfig}

\usepackage{booktabs}
\usepackage{array}
\usepackage{tabularx}   % 自适应列宽
\newcolumntype{Y}{>{\raggedright\arraybackslash}X} % 便捷可换行列

\usepackage{siunitx}
\usepackage{algorithm}
\usepackage{algpseudocode} % 或 algorithmic

\usepackage{tikz}
\usetikzlibrary{arrows.meta,positioning,shapes.geometric,fit}

\usepackage{xurl}                 % 允许 URL 在更多位置断行
\usepackage[hidelinks]{hyperref}  % 超链接（放靠后）

\usepackage[nameinlink,capitalize,noabbrev]{cleveref} % 放在 hyperref 之后

\usepackage{cite} % 数字压缩排序 [1]-[3] -> [1–3]，与 ieeeconf 兼容

\usepackage{amsmath,amssymb}
\usepackage{array,tabularx,booktabs}
\usepackage{float} % 为了 [H] 强制就地
\usepackage{array,makecell,tabularx,booktabs}
\newcommand{\yes}{\ensuremath{\checkmark}}
\newcommand{\no}{\textemdash}

\newcolumntype{L}{>{\raggedright\arraybackslash}p{.40\columnwidth}}
\newcolumntype{C}{>{\centering\arraybackslash}p{.10\columnwidth}}
\usepackage{xcolor}
\IEEEoverridecommandlockouts                              % This command is only needed if 
\begin{document}

\title{\LARGE \bf
Surgi-HDTMR: Closing the Sensorimotor Loop in Bimanual Microsurgery via Haptics, Digital Twin, and Mixed Reality
}

%\title{\LARGE \bf
%TriSens: Closing the Sensorimotor Loop in Bimanual Microsurgery via MR, Haptics, and Digital Twins}

\author{Songming Ping*, Shaoyue Wen*, Junhong Chen, Wen Fan, Lan Wei, Dandan Zhang
\thanks{*Equal contribution. All authors are with Imperial College London.}
}

\maketitle
\thispagestyle{empty}
\pagestyle{empty}

%%%%%%%%%%%%%%%%%%%%%%%%%%%%%%%%%%%%%%%%%%%%%%%%%%%%%%%%%%%%%%%%%%%%%%%%%%%%%%%%
\begin{abstract}
Robotic microsurgery demands precise bimanual control, intuitive interaction, and informative force feedback. However, most training platforms for robotic microsurgery lack immersive 3D interaction and high-fidelity haptics. Here, we present \textbf{Surgi-HDTMR}, a mixed-reality (MR) and digital-twin (DT) training system that couples bimanual haptic teleoperation with a benchtop microsurgical robotic platform, and 3D-printed phantoms. A metrically co-registered, time-synchronized DT aligns in-situ MR guidance with the physical workspace and drives a depth-adaptive haptic model that renders contact, puncture, and tissue-retraction forces. In a within-subjects study of simulated cortical navigation and tumor resection, \textbf{Surgi-HDTMR} shortened task time, reduced harmful contacts and collisions, and improved perceptual accuracy relative to non-haptic and non-adaptive baselines. These results suggest that tightly coupling MR overlays with a synchronized DT, together with depth-adaptive haptics, can accelerate skill acquisition and improve safety in robot-assisted microsurgery, pointing toward next-generation surgical training.

\end{abstract}

\section{INTRODUCTION}

Robot-assisted microsurgery (RAMS) is reshaping surgical care by delivering sub-millimeter precision, motion scaling, and tremor attenuation, enabling less invasive approaches~\cite{Zhang2022RAMSsurvey}. Early clinical studies, particularly in ophthalmology, report outcomes comparable to those of manual surgery, although operative times are often longer~\cite{Edwards2018NBE}. Its expanding use in neurosurgery, otolaryngology, and ophthalmology underscores both the opportunities and the risks of operating at the microscale, which highlight the need for high-fidelity microsurgical training~\cite{Lin2023Heliyon,zhang2020automatic,Maheo2023JCM}.

Despite this need, many microsurgical training programs face critical limitations~\cite{zhang2020microsurgical}. First, most systems rely on two-dimensional (2D) microscopic imaging, providing insufficient scene fidelity to replicate the spatial realism of the actual surgical workspace; trainees must infer 3D structure from flat projections, a mismatch that limits skill transfer in high-precision tasks~\cite{shuTwinSDigitalTwin2023}. Second, single-handed training paradigms remain dominant, limiting development of bimanual coordination required for simultaneous manipulation, retraction, and suturing~\cite{jiangDigitalTwinDrivenImmersive2024}. Third, the absence of force feedback deprives trainees of a critical guidance channel. Haptic cues convey tissue resistance and contact thresholds in real time, enabling precise force adjustment that vision alone cannot support. Without such feedback, mis-operations may remain undetected until clinical consequences occur~\cite{luApplicationsMixedReality2022}. Fourth, control is often fragmented across non–co-located modalities (foot pedals, keyboard) with poorly aligned visual guidance \cite{zhang2018self}, increasing cognitive workload and interrupting bimanual coordination.

\begin{figure*}[!tbp]
    \centering
    \captionsetup{font=footnotesize,labelsep=period}
    \includegraphics[width=0.84\textwidth]{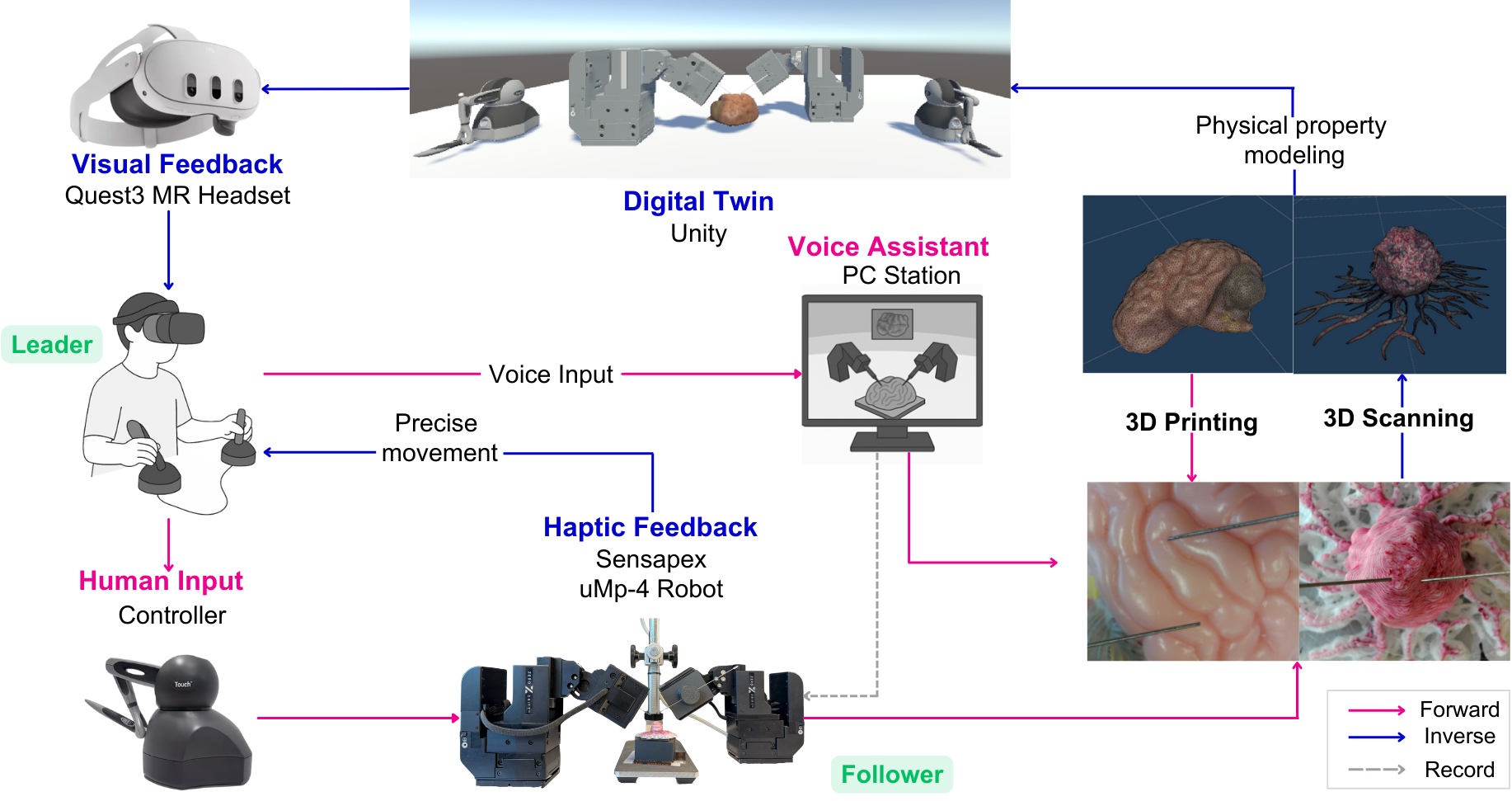}
    \caption{\textbf{System overview.} Mixed-reality (MR) surgical trainer in a leader–follower architecture. 
    The operator wears a Quest 3 head-mounted display (HMD) and controls two Touch controllers. 
    Hand increments are Kalman filtered and streamed to two Sensapex UMP-4 micromanipulators (\textit{forward loop}). 
    Ground-truth end-effector poses update the DT and drive depth-adaptive haptic rendering (\textit{reverse loop}). 
    Visual feedback (MR passthrough) closes the control loop.}
    \label{fig:system}
  \vspace{-0.6cm}
\end{figure*}

To address these gaps, we design an interface aligned with surgical practice: (i) co-registered mixed-reality (MR) overlays that align the visual-motor axis for direct, in-situ guidance; (ii) bimanual teleoperation with depth-responsive haptics that convey contact, puncture, and retraction cues. Prior work shows that MR improves 3D visualization and spatial awareness and is well accepted as a surgical aid~\cite{jiangDigitalTwinDrivenImmersive2024}. Restoring haptic sensation increases accuracy, limits excessive force, and can shorten completion time in robot-assisted tasks~\cite{okamura2004haptics,guo2024lightweight,bergholz2023haptic}.  Finally, tightly synchronized digital twins (DTs) couple virtual guidance to tangible manipulation, enabling adaptive feedback and objective performance assessment~\cite{asciak2025dtas,fan2023digital,ding2024sdsdt}.

As summarized in Table~\ref{tab:sota_cap_matrix}, prior systems typically cover only subsets of the required capabilities. Twin-S~\cite{shu2023twin} provides MR overlays with limited DT synchronization but lacks bimanual operation, and haptics. Shared-control DT frameworks~\cite{selvaggio2021shared} leverage a DT and virtual-fixture guidance rather than full haptic rendering. Networked DT teleoperation with UR3~\cite{kuts2022digital} demonstrates DT synchronization but omits immersive visualization and advanced interaction. VR-centric interfaces~\cite{greene2023dvpose,nguyen2024human} focus on visualization or camera control without bimanual haptic feedback. In contrast, Surgi-HDTMR uniquely integrates MR, depth-responsive haptics, and real-time DT synchronization for bimanual teleoperation as a comprehensive microsurgical training platform that addresses the gaps identified in prior work.

Here, we introduce a bimanual surgical training system, called \textbf{Surgi-HDTMR}. The system is designed around two core priorities: high-fidelity DT-driven spatial reconstruction and depth-responsive safety haptics, integrated with an MR interface. To support both functions without interference, we adopt a structurally decoupled multimodal architecture.
\begin{itemize}
  \item \textbf{MR interface:} Co-registered visual overlays align the visual-motor axis and project in-situ guidance directly onto physical phantoms, improving depth perception and situational awareness. Sub-millimeter metric consistency (1\,cm) is achieved through automated multi-view RGB-D fusion, providing depth cues absent in conventional 2D microscopic training environments.
  \item \textbf{High-fidelity DT with depth-adaptive haptics:} A metrically accurate, time-synchronized DT drives nonlinear force rendering that scales with tissue penetration depth. This provides contact, puncture, and retraction cues that (i) restore the proprioceptive feedback channel absent in vision-only systems and (ii) enforce proximity-aware safety constraints to prevent over-insertion near delicate anatomical structures.
  \item \textbf{Structurally decoupled multimodal architecture:} The DT reconstruction pipeline, haptic rendering pipeline, and MR interface operate independently while sharing a unified safety-bounded control bus. This design allows each modality to operate at optimal fidelity, supports modular ablation studies, and enables reuse across surgical procedures without re-engineering the control stack.
\end{itemize}

Through controlled user studies on representative microsurgical tasks (cortical surgery and tumor resection), we demonstrate that this integrated Surgi-HDTMR framework improves efficiency, precision, and user experience compared to conventional training approaches.
%------------------------------------

\begin{table}[H]
  \centering
  \scriptsize
  \captionsetup{font=footnotesize,labelsep=period}
  \caption{\textbf{Capabilities of representative systems.}}
  \label{tab:sota_cap_matrix}
  \resizebox{\columnwidth}{!}{%
  \begin{tabular}{l *{5}{c}}
    \toprule
    System (year) & VR & MR & Bimanual & Haptics & DT Sync \\
    \midrule
    \textbf{This work}      & \yes & \yes & \yes & \yes & \yes \\
    Twin-S\cite{shu2023twin}                & \no  & \yes & \no  & \no  & Partial \\
    Shared-control DT \cite{selvaggio2021shared}       & \yes & \no  & \no  & VF   & Uses DT \\
    UR3 DT over networks \cite{kuts2022digital}    & \no  & \no  & \no  & \no  & \yes \\
    dVRK HMD camera control \cite{greene2023dvpose} & \yes & \no  & \no  & \no  & \no \\
    VR teleop interface \cite{nguyen2024human}     & \yes & \no  & \no  & \no  & \no \\
    \bottomrule
  \end{tabular}%
  }
  \vspace{1pt}
  \par\noindent\scriptsize
  Legend: \yes{} provided; \no{} not reported; VF = virtual fixture guidance; “Partial/Uses~DT” indicates limited or indirect DT synchronization.
\end{table}

\section{Related Work}

\subsection{MR and DT for RAMS}
DT frameworks have been used to predict clinical outcomes (e.g., postoperative gradients)~\cite{golsePredictingRiskPosthepatectomy2021} and to create adaptable training models across medical contexts~\cite{katsoulakisDigitalTwinsHealth2024}. Within RAMS, DTs have supported low-latency teleoperation~\cite{tsokaloRemoteRobotControl2019} and shared control for novices~\cite{hagmannDigitalTwinApproach2021}. MR overlays with precise tracking (e.g., Twin-S) enhance in-situ situational awareness for guided manipulation~\cite{shu2023twin}. Networked DT synchronization studies (e.g., UR3 DT over networks) examine end-to-end state mirroring and latency effects, validating DT sync without full MR/VR or bimanual integration~\cite{kuts2022digital}. VR-centric camera interfaces for the dVRK improve viewpoint management but typically omit DT coupling and bimanual training behaviors~\cite{greene2023dvpose}; similarly, general VR teleoperation UI work emphasizes usability in single-instrument setups without MR overlays or DT sync~\cite{nguyen2024human}. Collectively, these efforts improve awareness and workload but often remain single-hand or task-specific and provide limited multimodal synchronization for complex \emph{bimanual} manipulation~\cite{jiangDigitalTwinDrivenImmersive2024}. 

\subsection{Haptics for RAMS}

\begin{comment}
Haptic feedback is central to restoring tactile cues and improving safety. Prior studies emphasize force reflection for precision control~\cite{patel2022haptic,garcia2024teleoperated}, while impedance-based guidance and virtual fixtures improve accuracy and reduce errors~\cite{malukhin2018mathematical,casini2015design}. Shared-control DT architectures implemented in VR demonstrate VF-style guidance but are commonly limited to single-instrument tasks without MR overlays or voice/NLP input~\cite{selvaggio2021shared}.     
\end{comment}

In microsurgical contexts, haptic rendering helps limit tissue strain and supports novice learning, with evidence of improved performance and lower error rates~\cite{lebrun2025mentor}. Recent systems combine multi-DoF haptics with DT-driven shared control for more natural, transparent interaction, yet challenges persist: maintaining stability under high-frequency updates, scaling across platforms, and achieving tight coupling with immersive MR environments~\cite{takahashi2024effects,jiang2024digital}. These gaps highlight the need to co-design haptics with MR/DT so that force cues remain stable, informative, and temporally aligned with visual guidance.

Crucially, while MR overlays have been adopted in surgical training platforms, no prior system simultaneously achieves: (1) automated high-precision DT reconstruction at metric scale co-registered with the physical workspace, and (2) depth-responsive haptic rendering that adapts force cues to instantaneous penetration depth in a bimanual teleoperation context. Existing MR platforms provide improved visualization but without the safety-critical proprioceptive channel; existing haptic platforms lack the spatial fidelity of a synchronized DT. Beyond these capability gaps, prior platforms also lack a structurally decoupled architecture: modalities are either absent or tightly fused into monolithic control policies, precluding independent validation and cross-task extension.

\section{Methods}\label{chap:methods}
\subsection{System Overview}
Our system integrates three complementary modalities:
(1) an immersive, DT-driven MR module that provides metrically accurate overlays and consistent depth cues, addressing the spatial misalignment and limited depth perception common in conventional VR simulators;
(2) a condition-adaptive haptic feedback module that restores tactile realism while enforcing proximity-based safety constraints to mitigate overshoot near delicate anatomy; and
(3) an intuitive human–robot interaction module (voice-command interface) that enhances operational transparency and control.
Together, these visual and haptic channels form the foundation of \textbf{Surgi-HDTMR}, a unified multimodal surgical training platform that combines spatial fidelity, safe haptic realism, and intuitive human–robot interaction (Fig.~\ref{fig:system}).
A key architectural principle is structural decoupling: the MR (pose data only), haptic (depth/velocity only), and DT construction pipeline are independent, converging on a shared safety-bounded control bus. Any channel can be enabled, disabled, or replaced without modifying the others.

% The system is supported by a leader–follower teleoperation framework, an automated 3D scene–reconstruction pipeline that produces metrically scaled DTs, and a low-latency networking layer, together preserving end-to-end fidelity from perception to actuation.
The system is supported by a leader–follower teleoperation framework, an automated 3D scene–reconstruction pipeline that produces metrically scaled DTs (matching real-world size and geometry), and a low-latency networking layer, together preserving end-to-end fidelity from perception to actuation.
Two \emph{Touch} haptic devices serve as leaders and drive dual Sensapex uMp-4 followers, with the scene rendered through a Quest~3 mixed-reality (MR) headset. A metrically co-registered DT maintains a fixed scale (\textbf{1\,cm $\leftrightarrow$ 0.1\,U}), supports both full-VR and passthrough-MR modes, streams Kalman-filtered hand-increment commands at 90\,Hz, and uses robot state feedback to update the twin and compute haptic forces in real time. A natural-language channel provides sterile, hands-free `nudges', protected by unified safety gates across modalities.

\subsection{MR for Teleoperation and Low-Latency Networking}
The MR module fuses the DT with the physical workspace, projecting overlays that are continuously corrected by robot ground truth to ensure robust spatial alignment during teleoperation.
At each update $i$, the follower pose is commanded as
\begin{equation}
\begin{aligned}
\mathbf{p}_f^{i}=\mathbf{p}_f^{i-1}+\alpha\,\mathbf{R}_{HW}\,\Delta\mathbf{p}_h^{i},
\end{aligned}
\label{eq:poseupdate}
\end{equation}

where $\Delta\mathbf{p}_h^{i}\!\in\!\mathbb{R}^3$ is the Kalman-filtered increment in the hand frame $H$, $\mathbf{R}_{HW}\!\in\!\mathrm{SO}(3)$ is the rotation into workspace $W$, and $\alpha>0$ is the motion-scaling factor~\cite{jiang2024adaptive}.
This incremental clutching maintains consistent one-to-one correspondence between the operator's hand and the micromanipulator in $W$.

\textbf{Haptic-to-Robot Frame Mapping and Calibration.} The system calibration establishes a fixed transformation between the haptic device frame $H$ and the robot workspace frame $W$. We use a rigid registration procedure based on physical fiducials: (i) the needle tip length (fixed at 35\,mm) provides the $z$-axis reference; (ii) the optical table dimensions (600$\times$400\,mm) establish the $xy$-plane; and (iii) a global isotropic scale of \textbf{1\,cm\,$\leftrightarrow$\,0.1\,Unity units} ensures metric consistency. The Sensapex uMp-4 micromanipulators provide sub-micron positioning accuracy (specified $<$\,10\,nm resolution, $<$\,1\,$\mu$m repeatability per manufacturer documentation~\cite{horan2024repix}), ensuring that the commanded increments are faithfully executed at the physical end-effector. 

Networking uses UDP with the latest-valid packet policy, discarding delayed or out-of-order messages. 
A watchdog bridges short dropouts via sample-and-hold, and extended silence beyond $T_{\mathrm{wd}}$ triggers safe-hold of the followers. 
Periodic clock re-synchronization bounds drift across devices and refreshes local timebases. 
End-to-end latency is maintained below a single 90\,Hz frame ($\approx 11.1$\,ms), preserving tight phase alignment among MR rendering, the haptic servo loop, and robot kinematics. 
This low-latency integration keeps visual overlays, haptic feedback, and teleoperation strictly time-synchronized, which is a requirement that many existing training systems do not satisfy.

\subsection{Digital Twin and Condition-Variable Haptic Rendering}
We reconstruct each surgical task as a metrically scaled DT: instrument geometries and operative workspaces are meshed from CAD or 3D scans (reality capture) and registered to the physical setup, while soft-tissue phantoms are fabricated by 3D printing flexible TPU to mirror gross anatomy and boundary conditions. The virtual and physical scenes share a common coordinate frame so that tool pose, contact state, and force feedback are consistent across modalities.

Each interactable object is assigned a \texttt{HapticMaterial} profile that parameterizes its interaction laws. These include stiffness and damping for normal contact, viscosity for rate-dependent bulk response, static and kinetic friction (optionally with Stribeck blending) for tangential contact, puncture thresholds with smooth hysteresis, and short-range adhesion. 
The profiles function as reusable presets that can be applied per surface region or tissue layer, enabling rapid switching among tissue types (e.g., fascia, parenchyma, vessel wall) or stratified compositions within a single organ. This modular parameterization supports the structural decoupling principle: haptic behavior is fully encapsulated within the material profile and operates independently of the MR subsystem, allowing haptic fidelity to be tuned without affecting visualization parameters.

Condition-variable rendering selects and blends these interaction laws based on measured or estimated state variables (penetration depth, relative velocity, contact history, distance-to-boundary, and tool identity). This structure supports predictable, passivity-friendly force updates at high servo rates while allowing fine-grained control over local material behavior and task-specific constraints.

Haptic rendering combines depth-adaptive impedance with tangential drag, extended with friction, puncture, and adhesion modules.
For penetration depth $d\!\geq\!0$, normal $\mathbf{n}$, velocity $\mathbf{v}$, and tangential component $\mathbf{v}_t$, the total force is
\begin{equation}
\begin{aligned}
\mathbf{F} ={}& (k_0+Ud)\,d\,\mathbf{n} - b(\mathbf{v}\!\cdot\!\mathbf{n})\,\mathbf{n}  - (c_0{+}c_1 d)\,\mathbf{v}_t \\
& {} - q_t\|\mathbf{v}_t\|\,\mathbf{v}_t - c_g\,\mathbf{v} 
+ \mathbf{F}_{\mathrm{fric}} + \mathbf{F}_{\mathrm{punc}} + \mathbf{F}_{\mathrm{extra}} .
\end{aligned}
\label{eq:force_total}
\end{equation}
\begin{comment}
\begin{equation}
\begin{aligned}
\mathbf{F} ={}& (k_0+Ud)\,d\,\mathbf{n} - b(\mathbf{v}\!\cdot\!\mathbf{n})\,\mathbf{n} \\[-2pt]
& {} - (c_0{+}c_1 d)\,\mathbf{v}_t - q_t\|\mathbf{v}_t\|\,\mathbf{v}_t - c_g\,\mathbf{v} \\[-2pt]
& {} + \mathbf{F}_{\mathrm{fric}} + \mathbf{F}_{\mathrm{punc}} + \mathbf{F}_{\mathrm{extra}} .
\end{aligned}
\label{eq:force_total}
\end{equation}    
\end{comment}

In Eq.~\ref{eq:force_total}: $k_0$ (N/mm) is the base stiffness; $U$ (N/mm$^2$) is the depth-dependent stiffness gain ensuring $k(d)=k_0+Ud$ increases monotonically with penetration; $b$ (N$\cdot$s/mm) is the normal damping coefficient; $c_0$ and $c_1$ parameterize depth-adaptive tangential drag; $q_t$ (N$\cdot$s$^2$/mm$^2$) is the quadratic velocity term for high-speed resistance; $c_g$ (N$\cdot$s/mm) provides global viscous damping for numerical stability. The function $U(d)$ is chosen as a linear ramp to ensure $C^0$ continuity at contact onset ($d=0$), with the gain tuned empirically to match soft-tissue stiffness ranges (0.1-1.0\,N/mm for brain parenchyma phantoms).

Normal impedance (stiffness and damping) increases monotonically with penetration depth, providing stronger restorative forces as contact deepens. The tangential drag adapts to instantaneous tangential velocity to capture shear-dependent effects. Puncture is modeled with a smooth, $C^1$-continuous hysteresis loop that separates loading and unloading branches, using a sigmoid transition function $\sigma(F_n - F_{\text{thresh}})$ where $F_{\text{thresh}}$ is the force threshold for membrane rupture. Upon puncture ($F_n > F_{\text{thresh}}$), stiffness temporarily reduces by 50\% for a brief window (50\,ms) to simulate membrane breakthrough, then recovers along the unloading branch. This avoids force discontinuities at rupture and during re-entry. %Proximity-based resistance channels generate anticipatory forces from distance-to-boundary estimates, acting as predictive safety margins that curb overshoot and minimize harmful dwell near constraints. 
All forces are explicitly bounded in magnitude and slew rate and are implemented with passivity-friendly coupling, yielding stable haptic rendering under high-frequency servo updates. These force profiles serve two purposes: they improve training realism by conveying tissue-specific proprioceptive cues, and they enhance safety by generating anticipatory resistance before insertion reaches damage thresholds, a function that vision alone cannot provide.

\textbf{Haptic Parameter Selection.} Parameters were tuned through iterative pilot sessions with two expert operators (one microsurgery trainee, one VR specialist) to achieve perceptually realistic tissue responses while maintaining device stability. Final values: $k_0=0.5$\,N/mm, $U=0.3$\,N/mm$^2$, $b=0.02$\,N$\cdot$s/mm, $F_{\text{max}}=3.3$\,N (device limit). Forces are updated at 1\,kHz and rendered at 90\,Hz to match the haptic servo loop, ensuring passivity under high-frequency contact transitions.

\subsection{Automated Scene Reconstruction and Digital Twin Generation}
To provide a metrically accurate basis for MR overlays and haptic rendering, Surgi-HDTMR incorporates an automated 3D scene reconstruction pipeline. 
A robotic arm executes a hemispherical sweep of calibrated viewpoints, capturing multi-view RGB data that is fused with depth streams into watertight meshes. 
High-resolution DSLR passes supply sharp textures, and all meshes are normalized such that 1\,cm corresponds to 0.1 Unity units. 
This automated pipeline eliminates parallax artifacts and guarantees metrically consistent, reproducible DTs (Fig.~\ref{fig:experiment}).

\section{User Study Design}
\subsection{Experimental Design and Procedure}
We conducted a within-subjects study to evaluate the effectiveness of \textbf{Surgi-HDTMR}. Before data collection, all participants completed a brief training session in the haptic force-feedback environment to standardize baseline familiarity with the system's interaction properties and its depth-responsive, nonlinear cues (gradual resistance, puncture-like transitions, and retraction forces). The training phase was not included in the analysis.

Two formal experimental tasks were conducted as shown in Fig.~\ref{fig:experiment}. The \textbf{brain needle tracing task} required participants to follow a predefined cortical trajectory using their dominant hand, emphasizing precision and stability for safe neurosurgical navigation. The \textbf{tumor resection task} demanded bimanual coordination: one hand stabilized the tumor model while the other followed the tumor boundary for continuous resection. At the midway point, hand roles were reversed to ensure each participant experienced both stabilizing and resecting operations.

Each participant completed both tasks under two system configurations: a \textbf{baseline condition} presenting only conventional 2D microscopic images through the monitor display, and the proposed \textbf{Surgi-HDTMR condition} enabling the full immersive framework with synchronized 3D DT visualization, depth-responsive haptic feedback. The counterbalanced within-subject design allowed direct comparison between baseline and Surgi-HDTMR modes, with three recorded trials per condition separated by short rest intervals following a Latin square schedule.

For quantitative analysis, we automatically recorded task efficiency (completion time), safety metrics (collision time and puncture events), kinematics (path length, velocities, movement variability), accuracy measures (supporting-hand distance to tumor apex), and system performance (frame rates). For qualitative analysis, participants completed NASA-TLX workload questionnaires \cite{noyes2007self} and semi-structured interviews focusing on MR visualization and DT force feedback after each condition.

\subsection{Participants}
We recruited 8 participants (5 male, 3 female; mean age =$26.3 \pm 3.2$ years) from medical schools with basic medical knowledge. Four had prior experience with MR systems, and three reported familiarity with robotic or microsurgical platforms, while one was a novice.  All participants had normal or corrected-to-normal vision and no history of neurological or musculoskeletal disorders. Informed consent was obtained under an IRB-approved protocol.

\begin{figure}[!t]
    \centering
    \captionsetup{font=footnotesize,labelsep=period}
    \includegraphics[width=1\linewidth]{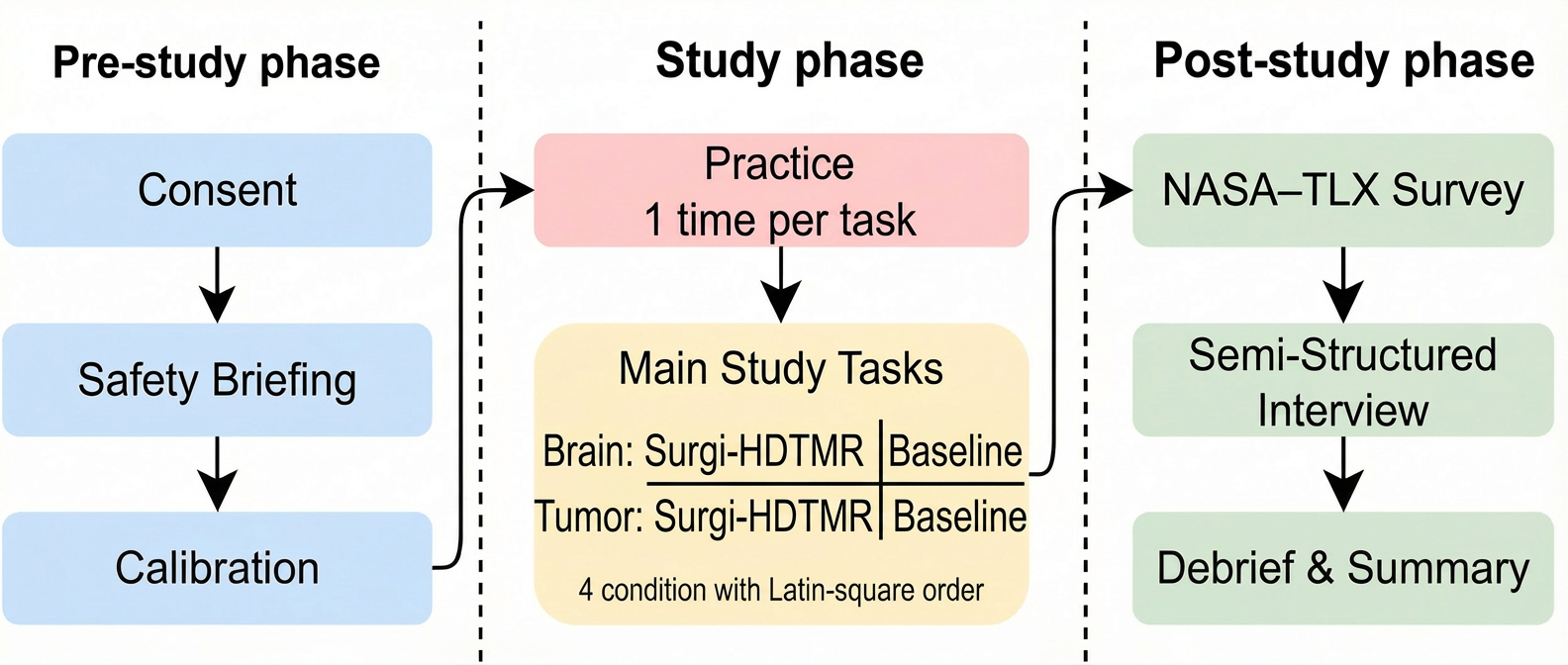}
\caption{Per-participant workflow: After consent and briefing, each participant completes one practice trial per task, followed by four counterbalanced blocks (Brain/Tumor × With/Without Surgi-HDTMR), with three recorded trials per condition and short rests. After each condition, NASA-TLX and a brief interview are completed. }
    \label{fig:experiment}
     \vspace{-0.4cm}
\end{figure}
   % \caption{Per-participant workflow. After consent and briefing, each participant practices once per task, then completes \textbf{four counterbalanced blocks} (Brain/Tumor $\times$ With/Without SurgiMR) with \textbf{three recorded trials per condition} and short rests. After each condition, NASA--TLX and a brief interview are completed. All trajectories/kinematics/events are recorded automatically.}

\begin{table*}[t]
  \centering
  \footnotesize
  \captionsetup{font=footnotesize,labelsep=period,aboveskip=2pt,belowskip=0pt}
  \caption{Metrics: definition, unit, and desired direction.}
  \label{tab:metric-summary}
  \renewcommand{\arraystretch}{1.1}
  \setlength{\tabcolsep}{5pt}
  \begin{tabularx}{0.85\textwidth}{l X c c}
    \toprule
    \textbf{Metric} & \textbf{Definition (brief)} & \textbf{Unit} & \textbf{Dir} \\
    \midrule
    Support-hand distance ($d_{\min}$, $\overline{d}$) & Support-hand distance to tumor apex (per hand/condition) & mm & $\downarrow$ \\
    $\Delta d^{(h)}$ & Mean distance gain (Orig $-$ Surgi-HDTMR) & mm & $\uparrow$ \\
    $L$ & Path length from world-mapped increments & mm & $\downarrow$ \\
    $T$, $\bar v$, $v_{\max}$ & Duration; avg/peak speed & s, mm/s & context$^\ast$ \\
    $\tau_{\text{coll}}$, $N_{\text{puncture}}$ & Collision time; puncture count & s, -- & $\downarrow$ \\
    $\min d_L,\min d_R$, $\rho_{<1\text{mm}}$ & Min clearance; sub-mm proximity ratio & mm, \% & safety$^\dagger$ \\
    FPS (avg/p1\%) & Runtime stability & Hz & $\uparrow$ \\
    \bottomrule
  \end{tabularx}
  \vspace{1pt}
  \par\noindent\tiny
  $\uparrow$ larger is better;\; $\downarrow$ smaller is better.\;
  $^\ast$ Faster if safety unchanged.\;
  $^\dagger$ Larger $\min d$ safer; lower $\rho_{<1\text{mm}}$ safer.
   \vspace{-0.2cm}
\end{table*}

\begin{figure*}[!t] % ← 改为 figure* 表示双栏通栏
    \centering
    \captionsetup{font=footnotesize,labelsep=period}
    \includegraphics[width=0.9\linewidth]{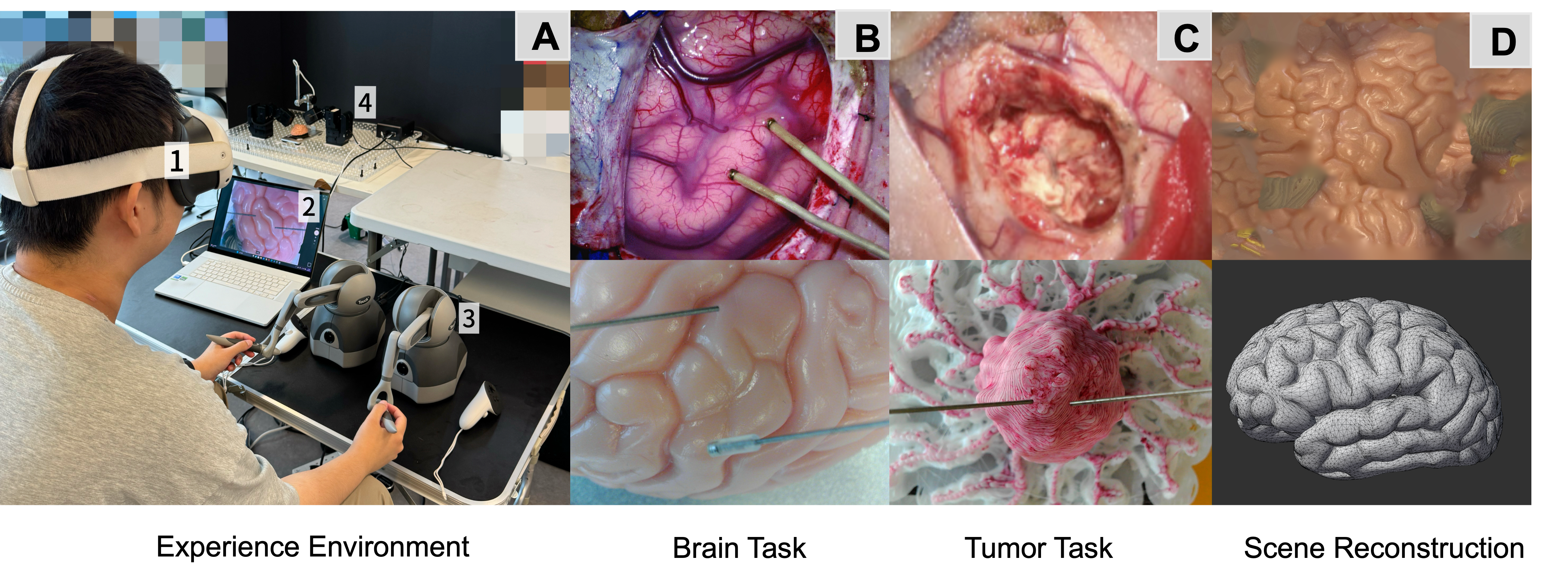}
      \vspace{-0.4cm}
    \caption{
(A) Experiment apparatus, including the MR head-mounted display (1), Unity-based PC visualization (2), dual haptic controllers (3), and the dual-arm Sensapex uMp-4 robotic platform (4).
(B) Needle tracing task, in which participants trace predefined cortical sulci using their dominant hand to assess precision and stability \cite{kim2018intraoperative}.
(C) Tumor resection task, requiring coordinated bimanual operation to stabilize and resect a tumor boundary, thereby simulating realistic microsurgical procedures \cite{stummer2017fluorescence}.
(D) 3D Scene Reconstruction, showing how volumetric cortical and tumor geometries are reconstructed and rendered in MR for immersive interaction.}
    \label{fig:experiment}
      \vspace{-0.2cm}
\end{figure*}
%---------------------------------------------------

%-----------------------------------------------
\section{Results}
\subsection{Quantitative Results}
Given the small sample size and non-normal distributions, we applied a Friedman test for omnibus comparisons across the four conditions~\cite{friedman1937}. When the omnibus was significant or marginal, we performed Wilcoxon signed-rank tests to compare \textit{Surgi-HDTMR} vs.\ \textit{baseline} within each task (Brain/Tumor)~\cite{wilcoxon1945}. Two-sided tests were used throughout, with Holm correction applied across follow-ups~\cite{holm1979}.

\begin{figure}[t]
  \centering
  \captionsetup{font=footnotesize,labelsep=period}
  \subfloat[Brain task - with \textit{Surgi-HDTMR}\label{fig:brain_with_surgimr}]{%
    \includegraphics[width=0.48\columnwidth]{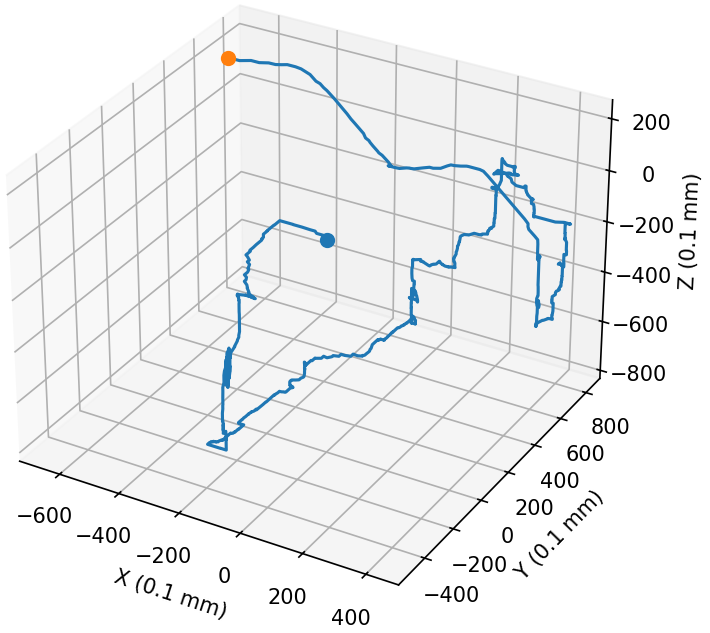}%
  }\hfill
  \subfloat[Brain task - without \textit{Surgi-HDTMR}\label{fig:brain_without_surgimr}]{%
    \includegraphics[width=0.48\columnwidth]{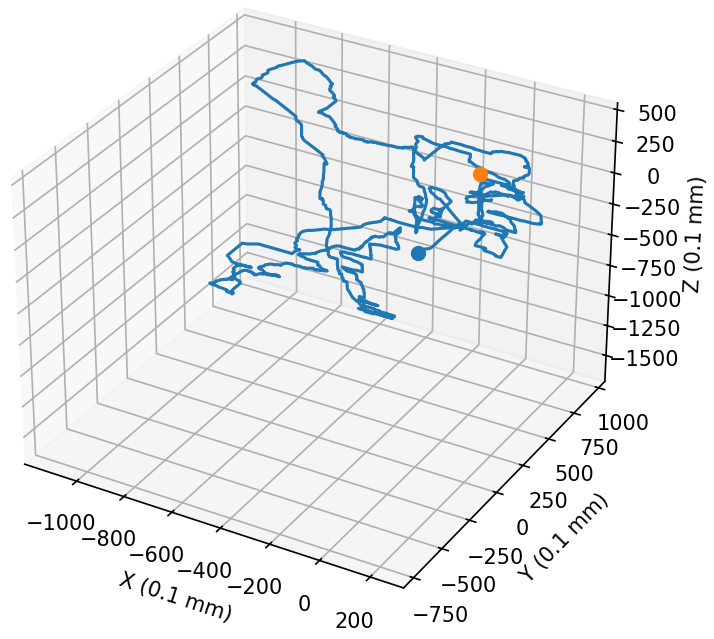}%
  }
  \caption{Bimanual end-effector trajectories for the brain task with and without \textit{Surgi-HDTMR}.
The framework promotes early acquisition of role-differentiated bimanual strategies in microsurgery and improves coordination between stabilizing and manipulating hands, which traditional single-hand or vision-only platforms often fail to develop effectively.}
  \label{fig:brain_trajectories_surgimr}

\end{figure}

\begin{figure*}[htbp]
    \centering
  \vspace{-0.6cm}
    \captionsetup{font=footnotesize,labelsep=period}
    \includegraphics[width=0.8\linewidth]{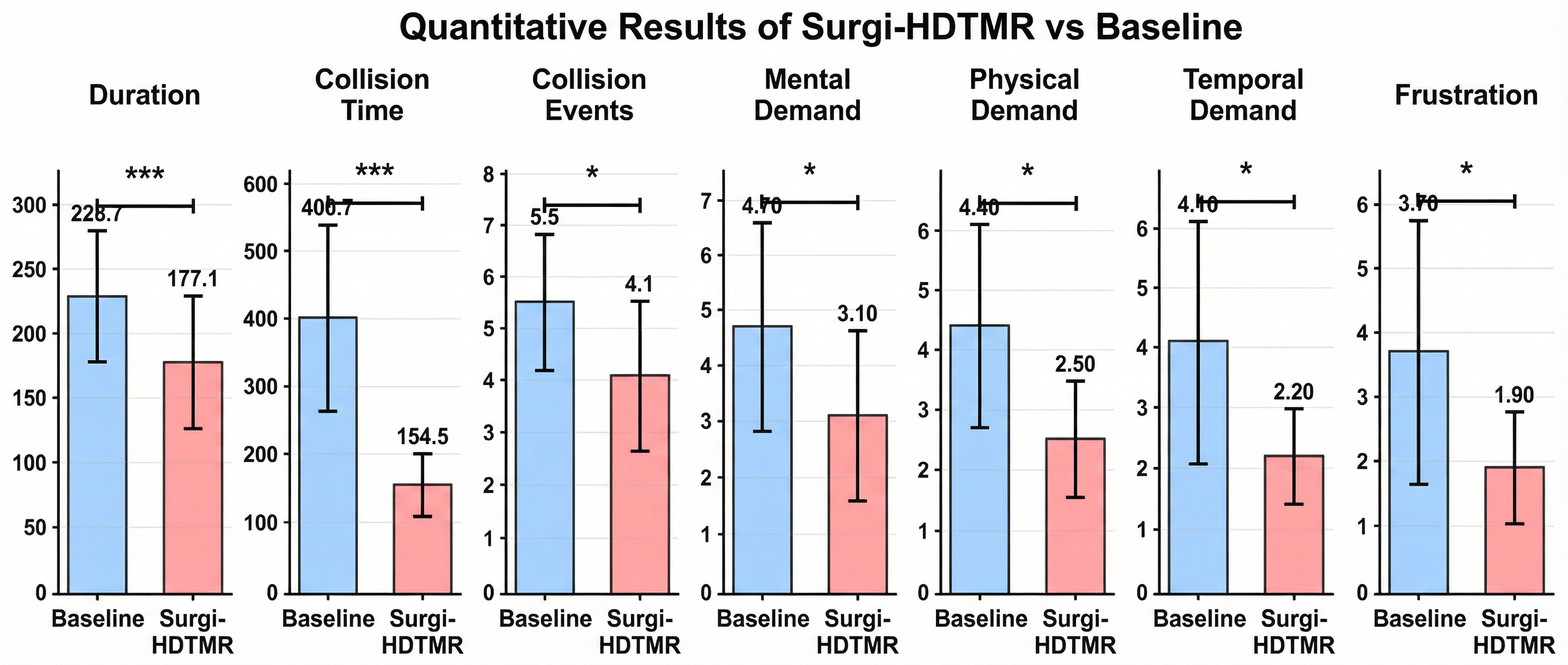}
    \caption{Performance and workload comparison between Surgi-HDTMR and baseline systems. Significant differences ($p < 0.05$) shown for task efficiency metrics (duration, collision time, collision events) and NASA-TLX subjective workload measures (mental demand, physical demand, temporal demand, frustration). Error bars represent standard error. Statistical significance: *$p < 0.05$, **$p < 0.01$, ***$p < 0.001$.}
    \label{fig:boxplots_6in12}
    \vspace{-0.4cm}
\end{figure*}

\textbf{Task efficiency} was measured as completion time (seconds) required to complete each task, calculated from task initiation to completion. Results showed significant differences across conditions, $\chi^2(3){=}12.20$, $p{=}0.007$. Wilcoxon tests indicated shorter durations with Surgi-HDTMR for both Brain ($Z{=}{-}2.201$, $p{=}0.028$) and Tumor ($Z{=}{-}2.201$, $p{=}0.028$) tasks. The identical signed-rank outcomes reflect matching within-subject rank patterns in this sample.

\textbf{Collision time} was quantified as cumulative duration (seconds) of end-effector contact with tumor/phantom tissue, indicating potential tissue damage risk. A significant omnibus effect was observed, $\chi^2(3){=}15.20$, $p{=}0.002$. Post hoc tests showed reduced collision duration with Surgi-HDTMR in both Brain ($Z{=}{-}2.201$, $p{=}0.028$) and Tumor ($Z{=}{-}2.201$, $p{=}0.028$) tasks. As above, identical test statistics arise from the same rank structure across paired samples.

\textbf{Collision events} were measured as the number of discrete puncture events that captured excessive insertions potentially damaging tissue. Results differed overall, $\chi^2(3){=}10.29$, $p{=}0.016$, with fewer punctures using Surgi-HDTMR in the Brain task ($Z{=}{-}2.214$, $p{=}0.027$) but no difference in the Tumor task ($Z{=}{-}0.412$, $p{=}0.680$), suggesting task-dependent safety improvements.

\textbf{Movement stability} was assessed through speed variability, calculated as the standard deviation of velocity (mm/s) during task execution. Results yielded a marginal omnibus effect, $\chi^2(3){=}7.20$, $p{=}0.066$. In the Tumor task, Surgi-HDTMR significantly reduced variability ($Z{=}{-}1.992$, $p{=}0.046$), whereas the Brain task showed no difference ($Z{=}{-}1.363$, $p{=}0.173$), suggesting improved movement stability with our system in the more demanding bimanual scenario.

\textbf{Accuracy} was measured as supporting-hand distance to tumor apex (mm), calculated as the Euclidean distance between the supporting-hand tool tip and tumor center, indicating fixation precision. Wilcoxon tests revealed significant improvement with Surgi-HDTMR for the left hand ($Z{=}{-}2.207$, $p{=}0.027$) and a trend toward improvement for the right hand ($Z{=}{-}1.732$, $p{=}0.083$), suggesting enhanced anchoring stability, particularly for the non-dominant side, compared to baseline.

% preamble:
% \usepackage{makecell,tabularx,booktabs}

% preamble:
% \usepackage{array,makecell,tabularx,booktabs}

Several metrics showed no significant differences between Surgi-HDTMR and baseline. Path length, calculated as total trajectory length (mm) by summing frame-wise displacements, showed no significant differences across conditions ($\chi^2(3){=}5.00$, $p{=}0.172$), with non-significant pairwise comparisons (Brain: $Z{=}{-}0.105$, $p{=}0.917$; Tumor: $Z{=}{-}0.943$, $p{=}0.345$). Average speed (mm/s), computed as mean velocity throughout task execution, showed only a marginal omnibus effect ($\chi^2(3){=}7.20$, $p{=}0.066$) with non-significant pairwise comparisons (Brain: $Z{=}{-}1.363$, $p{=}0.173$; Tumor: $Z{=}{-}1.572$, $p{=}0.116$). Although these two values match those of the speed-variability analysis, they correspond to distinct metrics and resulted from the same participant-level rank ordering.

\textbf{Component Contribution Analysis.} While the study compared the full Surgi-HDTMR system against a non-haptic 2D baseline, the pattern of results provides insight into component contributions: (i) \textbf{High-fidelity DT reconstruction} enhanced spatial realism through MR visualization, accelerating task completion by providing direct depth cues without increasing path length; (ii) \textbf{Depth-responsive haptics} served as a real-time guidance channel, contributing to reduced collision time and fewer punctures by signaling contact states and tissue resistance that vision alone cannot convey; (iii) \textbf{DT synchronization} enabled the smooth, low-latency feedback loop reflected in reduced speed variability, particularly in the bimanual Tumor task (Fig.4).

% 需要 \usepackage{tabularx,booktabs,array}

% 需要 \usepackage{tabularx,booktabs,array}
\vspace{4pt}
\begin{table*}[t] % 双栏表格
  \centering
  \scriptsize
  \captionsetup{font=footnotesize,labelsep=period,aboveskip=2pt,belowskip=0pt}
  \setlength{\tabcolsep}{5pt}
  \renewcommand{\arraystretch}{1.05}
  \caption[Inferential statistics across outcomes]{Inferential statistics across primary outcomes (Friedman omnibus and Wilcoxon follow-ups).
   Direction indicates the change under \emph{with} relative to \emph{without}.Holm correction applied across Brain or Tumor follow-ups.}
  \label{tab:inferential_summary}

  \begin{tabularx}{0.8\textwidth}{l c c c c >{\raggedright\arraybackslash}X}
    \toprule
    Metric & Friedman $\chi^2(3)$ & $p$ & Brain: $Z$/\,$p$ & Tumor: $Z$/\,$p$ & Direction (\projectname) \\
    \midrule
    Duration           & 12.20 & \textbf{0.007} & \textbf{$-2.201$/0.028}  & \textbf{$-2.201$/0.028}  & Shorter \\
    Path Length        & 5.00  & 0.172          & $-0.105$/0.917           & $-0.943$/0.345           & No change \\
    Average Speed      & 7.20  & 0.066          & $-1.363$/0.173           & $-1.572$/0.116           & Slight increase (n.s.) \\
    Speed SD           & 7.20  & 0.066          & $-1.363$/0.173           & \textbf{$-1.992$/0.046}  & More stable (Tumor) \\
    Collision Time     & 15.20 & \textbf{0.002} & \textbf{$-2.201$/0.028}  & \textbf{$-2.201$/0.028}  & Shorter \\
    Collision Events   & 10.29 & \textbf{0.016} & \textbf{$-2.214$/0.027}  & $-0.412$/0.680           & Fewer (Brain) \\
    Accuracy (anchor, Left hand)  & 13.94 & \textbf{0.003} & \textbf{$-2.207$/0.027} & --- & More precise (Left) \\
    Accuracy (anchor, Right hand) & 13.94 & \textbf{0.003} & --- & $-1.732$/0.083 & Trend (Right) \\
    \bottomrule
  \end{tabularx}
        \vspace{-0.4cm}
\end{table*}

\subsection{Qualitative Results}
Post-session interviews were conducted with all participants to capture their subjective experiences with the core innovations: Mixed Reality visualization and force-feedback digital twin system. NASA-TLX assessments showed consistent workload reduction with Surgi-HDTMR (Fig.~\ref{fig:boxplots_6in12}), complementing our quantitative findings of improved efficiency and reduced collision time. The Brain task showed reduced Temporal Demand (4.10→2.20, p=0.066) and improved Mental/Physical Demand. The Tumor task demonstrated significant reductions in Mental Demand (5.0→3.4, p=0.041) and Frustration (3.70→1.90, p=0.046), aligning with our quantitative results showing improved movement stability and accuracy in the more complex bimanual scenario.

Thematic analysis identified five key themes (Table~\ref{tab:qualitative_themes}) directly related to our core contributions. Participants particularly valued the MR's enhanced spatial perception, which correlated with our quantitative findings of improved accuracy and reduced collision events. The DT's realistic force feedback was consistently praised for providing intuitive depth cues as a real-time guidance channel, supporting our observed reductions in collision time across both tasks.

\begin{table}[t]
\centering
\scriptsize
\setlength{\tabcolsep}{2.5pt}
\renewcommand{\arraystretch}{1.02}
\captionsetup{font=footnotesize,labelsep=period,aboveskip=2pt,belowskip=0pt}
\caption{Qualitative themes with representative quotes.}
\label{tab:qualitative_themes}
% m{..} centers vertically; \centering centers horizontally in the first column
\begin{tabularx}{\columnwidth}{@{}>{\centering\arraybackslash}m{20mm}@{\hspace{0.5pt}}X X@{}}
\toprule
\textbf{Subtheme} & \textbf{Summary} & \textbf{Representative quotes} \\
\midrule
\makecell{MR\\immersion} &
MR overlays improved depth/spatial awareness; aligned with fewer collisions &
P2: \emph{"3D needle line helped judge contact"}; P1: \emph{"More intuitive than 2D"} \\
\makecell{DT force\\feedback} &
Depth-responsive resistance signaled contact/puncture &
P2: \emph{"Too deep-resistance jumped"}; P3: \emph{"Clear tip contact"} \\

\makecell{System\\responsiveness} &
Stable FPS and DT–MR sync improved smoothness &
P1: \emph{"Higher frame rate"}; P4: \emph{"Much smoother"} \\
\makecell{Anatomical\\fidelity} &
High-res DT/phantoms increased realism &
P2: \emph{"Details clear-near 1:1"}; P1: \emph{"Almost like real surgery"} \\
\makecell{Training\\feasibility} &
Integrated MR+haptics seen as useful for education &
P3: \emph{"Feasible for students"}; P2: \emph{"Safe practice before surgery"} \\
\bottomrule
\end{tabularx}
\vspace{-0.4cm}
\end{table}

\section{Discussion}
\label{sec:discussion}

\subsection{Stability over Speed: The Critical Performance Metric}
Our findings demonstrate that the fundamental value of the Surgi-HDTMR bimanual training system lies not only in accelerated task completion, but in enhanced control stability during critical tissue interactions. Despite equivalent path lengths and average velocities between conditions, participants exhibited significantly reduced exposure time in hazardous contact zones and more consistent supporting hand anchoring under the Surgi-HDTMR framework. These results underscore a fundamental principle of microsurgical proficiency: expertise is characterized not by rapid movement execution, but by the ability to maintain precise, deliberate control under uncertainty.

The depth-responsive haptic feedback and synchronized DT visualization collectively foster what we term \textit{protective precision}, a training paradigm that prioritizes safety margins over task efficiency. This observation suggests that conventional speed-based metrics may not fully capture the fine control strategies required for microsurgical proficiency. The structurally decoupled architecture further enables principled ablation: any single modality can be disabled without modifying the others. 

\subsection{Educational Transformation}
From a pedagogical perspective, the Surgi-HDTMR framework facilitates early acquisition of role-differentiated bimanual strategies fundamental to microsurgical practice. Participants demonstrated improved coordination between the stabilizing and manipulating hands, a critical skill set that traditional single-hand or vision-only training platforms often fail to cultivate effectively. The integration of immersive visualization and near-field haptic cues creates an interactive learning environment that extends beyond the limitations of conventional simulation approaches.

% two side-by-side trajectory maps (brain task)

These systems function not merely as training simulators, but as intelligent pedagogical companions that scaffold safe behavioral patterns while reducing extraneous cognitive load. This paradigm shift enables instructors to focus on higher-order clinical decision-making and strategic thinking rather than basic motor skill acquisition, fundamentally restructuring the educational trajectory from novice to competent practitioner.

\subsection{Limitations and Future Work}

To further improve realism and translational relevance, we plan to incorporate closed-loop perception that fuses vision and force for online estimation of contact state and tissue compliance, enabling adaptive virtual fixtures and safety constraints. Interaction modalities will be extended with gaze and context-aware speech, accompanied by systematic studies of modality arbitration and cognitive load. In addition, baseline conditions will be refined to more closely replicate traditional operative microscope viewing, supporting fairer comparisons with established training paradigms. 
Although experiment sessions with two expert microsurgical trainees informed parameter tuning (e.g., haptic gain thresholds) and confirmed perceptual plausibility relative to cadaveric experience, the present findings should be interpreted as demonstrating system feasibility rather than providing definitive evidence of training efficacy validated through clinical comparison. Larger-scale and longitudinal studies will therefore be conducted, including collaborative and remote training scenarios via a shared online platform.

\section{Conclusions}

We presented \textbf{Surgi-HDTMR}, a bimanual microsurgical training system that improves task efficiency, safety, and precision compared with conventional 2D training baselines. The system employs a structurally decoupled architecture in which high-precision DT reconstruction and depth-adaptive haptic rendering operate as independent pipelines on a shared safety-bounded control bus. This design enables simultaneous spatial fidelity and real-time haptic guidance without mutual interference, while supporting modular ablation and adaptation across surgical procedures. Future work will expand validation to practicing surgeons and surgical residents to evaluate clinical training transfer, and extend task coverage to ophthalmic and ENT microsurgical procedures using procedure-specific phantoms.

\bibliographystyle{ieeetr}
\bibliography{reference}

\end{document}